\documentclass{aastex}

\received{}
\revised{}
\accepted{}

\shorttitle{the X-ray Forest}
\shortauthors{Fang \& Canizares}

\begin{document}

\title{Probing Cosmology with the X-ray Forest}

\author{Taotao Fang AND Claude R.Canizares}
\affil{Department of Physics and Center for Space Research}
\affil{Massachusetts Institute of Technology}
\affil{Room 37-624D, 70 Vassar Street, Cambridge, MA 02139}
\email{fangt@space.mit.edu, crc@space.mit.edu}

\begin{abstract}

There is a growing consensus that in  the
present universe most baryons reside in galaxy clusters and
groups in the form of highly ionized gas at temperatures of $10^{6}
\sim 10^{8}$ K. The H-like and He-like ions of the heavy elements
can produce absorption features - the so-called ``X-ray Forest'' -
in the X-ray spectrum of a background quasar. We investigate the
distribution of the X-ray absorption lines produced by this gas under
three different cosmological models: the standard CDM with
$\Omega_{0}=1$, a flat model with $\Omega_{0}=0.3$ and an open model
with with $\Omega_{0}=0.3$. We give a semi-analytic calculation of the
X-ray forest distribution based on the Press-Schechter formalism,
following \citet{PL98}. We choose three ions (\ion{O}{8}, \ion{Si}{14}
and \ion{Fe}{25}) and calculate the distribution functions, the number of
absorbers along the line-of-sight (LOS) to a distant quasar
vs. redshift and column density in a given ion. We find that significant
differences in the evolution of the distribution functions among the three
cosmological models. Using Monte Carlo simulations, we simulate the
distribution of X-ray absorption lines for 10,000 random LOS. We find
there are at least several \ion{O}{8} lines with column density higher
than $10^{16} {\rm cm^{-2}}$. Finally we explore the possibility of detecting
the X-ray forest with current and upcoming X-ray missions and we
present an {\sl XMM}
RGS simulation of a representative quasar X-ray spectrum.

\end{abstract}


\section{Introduction}

There is an apparent deficit in the total density of baryons at
moderate  and low redshift, $z$, which has come to be called the
``missing baryon''  problem (see, e.g., \citealp{FHP98, CO99}).
Observations of the hydrogen and helium absorption lines in the Lyman
alpha forest give a baryon density at high redshift (z $\sim$ 3) of
$\Omega_{b} \geq 0.017h^{-2}$ \citep{RMS98}, which is consistent with the value  $\Omega_{b} =
0.019h^{-2}$ derived from standard big bang nucleosynthesis
\citep[a Hubble constant of $H_{0} =
100h$ km ${\rm s}^{-1}$ ${\rm Mpc}^{-1}$ is used throughout the
paper]{BT98}. However, in the local universe the baryon budget is far below
this number.  Summing over all the baryons inside stars, neutral
atomic gas and molecular gas gives $\Omega_{b} \sim 0.003h^{-1}$ .

There is a growing consensus that these ``missing baryons'' reside in
a hot ($\geq 10^6$ K), ionized plasma associated with groups of
galaxies.  This is effectively attributing to the universe a mix of
components similar to that observed in richer groups and galaxy
clusters (Fukugita et al. 1998; \citealp{CO99}). It is plausible that this medium can
be enriched in heavy elements. These heavy elememts would not be fully
ionized at the temperatures of interest, as is the cluster gas \citep{R97}.

Resonant absorption by a hot, enriched medium would introduce features
in the X-ray spectrum of a distant quasar. The features would be
narrow lines or broad troughs depending on the velocity structure of
the absorber.  The associated absorption edges would also be present
but are generally much weaker \citep{M99}. Early
work by  \citet{SB80} discusses the X-ray
absorption spectrum introduced by a uniformly distributed, hot IGM
with an admixture of metal atoms via X-ray ``Gunn-Peterson''
effect. Using the same method but with a photoionized model of the IGM,
\citet{AEM94} constrain the density and
temperature of the IGM with {\sl ROSAT} PSPC spectra of  $z \sim 3$
quasars. \citet*{BKM81} discusses the
detectability of X-ray resonance absorption lines in quasar spectra
produced by hot plasma in an intervening galaxy cluster. More detailed
and accurate calculations of cluster absorption were performed by
\citet{GSC87, KR88, S89}.

Like the Lyman alpha forest system in the optical band, the spatial
distribution of the galaxy groups and clusters can also produce an
``X-ray forest'' along the line of sight in the X-ray spectrum of a
background quasar. The concept of an ``X-ray forest'' was first suggested
by \citet{HGM98}: X-ray absorption lines are
produced by hot intergalactic medium in the form of ``filamentary and
sheet-like structures connected to galaxy clusters and groups, as well
as colder gas left out in voids.'' A similar concept was explored by
\citet{PL98} using the expected spectrum of mass
concentrations for a universe dominated by cold dark matter. The
effect of differing cosmologies on the number and evolution of X-ray
absorption lines is related to the effect on the number density of
clusters \citep{BF98,ECF98}. This means that X-ray absorption
line studies might eventually provide independent constraints on
cosmological parameters.

Here we build on and extend the approach of \citep{PL98} to explore the X-ray forest for various cosmologies and to
assess its detectability.  Detecting it is not easy.  Most of the
absorption lines will have equivalent widths $\leq 1$ eV, and none of
the previous X-ray missions ({\sl Einstein, ASCA, ROSAT})  had
sufficient sensitivity to detect such features in a quasar spectrum.
However, current and future missions give order-of-magnitude advances
in sensitivity for the X-ray forest. For example, {\sl Chandra}
grating spectrometers have resolving powers of 1,000 around 1 keV
\citep{ASC97} with sufficient collecting
area to detect an absorption line from  an ion column density of
approximately $10^{16} {\rm cm}^{-2}$ in a plausible quasar spectrum.
Assuming a moderate metal abundance, this column density implies a
hydrogen column density of a  modest galaxy cluster
\citep{CF98}. {\sl XMM} and {\sl
Constellation-X} achieve comparable or better energy resolutions with
larger effective area. In this paper we use simulations to assess the
ability of these missions to detect  features in the X-ray forest.

This paper is organized as follows: section II gives a semi-analytic
calculation of the X-ray forest distribution function, based on
Press-Schechter formalism. Section III is devoted to the numerical
simulation of the distribution function. In section IV we discuss the
detectability of the X-ray forest. Section V presents the conclusions and
discussion.

\section{X-ray Forest Distribution Function}

\subsection{Press-Schechter Formalism}

Galaxies, galaxy clusters and other large scale structures grow from
the initial small scale density fluctuation via gravitional
instability. The small scale fluctuation first grows linearly, until
it reaches a  critical density. Then it decouples from the Hubble
expansion, starts collapsing and finally condenses out as a
virialized, gravitational bound halo. Given a random Gaussian
distribution, the comoving number density of virialized halos can be
described by the Press-Schechter function :
\begin{equation}
\frac{dn}{dM_{vir}} =
  \left(\frac{2}{\pi}\right)^{\frac{1}{2}}\frac{\bar{\rho}}{M_{vir}^{2}}\frac{\delta_{c}}{\sigma(z,
  M_{vir})}\left|\frac{d\ln\sigma}{d\ln
  M_{vir}}\right|exp\left(-\frac{\delta_{c}^{2}}{2\sigma(z,
  M_{vir})^{2}}\right)
\end{equation}Here $M_{vir}$ is the mass of the virialized halo; $\bar{\rho}
  \equiv 3\Omega_{0}H_{0}^{2}/8\pi G$ is the comoving mean  density of
the universe which is constant during matter domination; $\delta_{c}$
denotes the linearly extroplated overdensity at which an object
virializes; and $\sigma(z, M_{vir})$ is the rms density fluctuation
inside halos containing a mean mass of $M_{vir}$. \citet{PS74} first gave this function with a very simple and
intuitive model : large scale virialized objects form from the
nonlinear interaction of small scale objects through a self-similar
condensation process. However, the original theory suffers the
so-called ``cloud-in-cloud'' problem of miscounting the underdense
regions properly. \citet{BCE91} and
\citet{LC93} extended this model by counting the
overdense regions one-by-one and gave the correct normalization of the
mass function. Although no direct observational evidence shows that
the Press-Schechter function is the right way to describe the cluster
abundance, this function fits N-body simulations extremely well \citep{LC94,ECF96,T98,F99}.

Formally one would expect $\delta_{c}$ to depend on the cosmological
model and the geometry of the collapsing object. Since most of the
rich clusters are fairly round it would be a good assumption that the
collapse is close to spherical. In a flat universe spherical collapse
gives $\delta_{c} = 1.686$ \citep{LC93}. However, the value of $\delta_{c}$ changes by $\leq 5\%$ as one
goes from an Einstein-de Sitter universe to an $\Omega = 0.1$ universe
\citep{ECF96}. So we adopt the value of 1.686
throughout the paper.

The rms mass variance at redshift $z$
can be  expressed by the present rms mass variance and the linear
growth factor $D(z)$ through\begin{equation} \sigma(z, M) = \sigma(z=0,
M)\frac{D(z)}{D(z=0)}
\end{equation}
From observations of cluster density in the local
universe, several papers give the normalization of the power spectrum
at $8h^{-1}{\rm Mpc}$ scale \citep{ECF96, VL96,P98, VL99}. Here we
adopt the value from \citet{ECF96}. The present rms mass fluctuation
$\sigma(z=0, M)$ can be calculated by the normalized power spectrum. A
functional fit to $D(z)$ is given by \citet{LRL91} 

To apply the Press-Schechter formalism we need to determine the virial
mass of clusters ($M_{vir}$) precisely. However, measuring the virial
mass of clusters is difficult, especially at high
redshift. Observationally the X-ray temperature ($T_{X}$) of galaxy
clusters can be measured more accurately. Numerical simulations and
observations of X-ray clusters show that there exists a very tight
relationship between $T_{X}$ and $M_{vir}$ \citep{HOK98, BN98}. To
simply we assume a singular isothermal sphere model of the cluster
mass density distribution. The mass-temperature relationship is given
by \citet{ECF96}
\begin{equation}
   kT_{X} =
   \frac{1.39}{\beta}\left(\frac{M_{vir}}{10^{15}h^{-1}M_{\odot}}\right)^{\frac{2}{3}}(1+z)\left(\Delta_{c}\frac{\Omega_{0}}{\Omega(z)}\right)^{\frac{1}{3}}
   {\rm keV}
\end{equation} Here $\Delta_{c}$ is the ratio of the mean cluster
density to the critical density at that redshift, $\beta$ is the ratio
of the specific kinetic energy to thermal energy, $\Omega_{0}$ and
$\Omega(z)$ are the cosmology density parameter at present and
redshift of $z$, respectively.

Recently various authors show that equation (3) is accurately obeyed
in N-body hydrodynamic simulations with value of $\beta \simeq 1$
\citep{NFW95,EMN96,BN98}. Although all of
them suggest a slightly higher $\beta$-value, it might be due to the
incomplete thermalization of the intracluster gas, or the gas density
dropping faster than $r^{-2}$ around the virial radius in the
numerical simulations. Here we use $\beta = 1$, which means that the
specific galaxy kinetic energy equals the specific gas thermal energy
within the virial radius. A Recent
analysis on Abell 401 shows the cluster mass given by the best-fit
model is approximately by a factor of 1.7 lower than the value
predicted by equation (3) \citep{NMF99}. The difference is attributed
to the fact that the simulated clusters
have steeper gas density and shallower temperature profiles than
observed.

\subsection{Gas Column Density Profile within Galaxy Clusters}

Assuming a ``$\beta$ model'' of the cluster gas density distribution
\citep{S88}, the column density of gas particles at an impact distance
of $b$ is \citep{PL98}
\begin{equation}
N_{gas}(b) = \frac{f_{gas}kT_{X}}{2Gr_{c}(\mu
m_{p})^{2}}\left[1+\left(\frac{b}{r_{c}}\right)^{2}\right]^{-\frac{1}{2}}
\end{equation}Here $b$ is the projected distance from the center of
the galaxy cluster; $r_{c}$ is the core radius of the galaxy cluster
and we select a constant value of 250 kpc throughout the paper;
$f_{gas}$ is the baryonic gas fraction; $\mu = 0.59$ is the mean
atomic weight and $k$ is the Boltzmann constant.

We are interested in the metal ion column density  which would
produced absorption features in the X-ray spectrum of a background
quasar. The ionization sources of intracluster gas can be either
photoionization from the X-ray background radiation or collisional
ionization. However the X-ray background is in general too weak to be
the main source of ionization so we only consider collisional
ionization here. Since generally the collisional time scale is much
shorter than Hubble time, the gas is in collisional ionization
equilibrium. If we denote $\Upsilon(X^{i}) \equiv N(X^{i})/N(X)$ as
the fraction of ion $X^{i}$, $\Upsilon$ would be only a function of
temperature, i.e., $\Upsilon = \Upsilon(T)$ under collisional
equilibrium \citep{SB77}. Assuming a uniform
metallicity $Z(X) \equiv N(X)/N(H)$ the ion $X^{i}$ column density
distribution is
\begin{equation}
N(X^{i}) = 0.46\ Z(X)\Upsilon(T)\frac{f_{gas}kT_{X}}{2Gr_{c}(\mu
  m_{p})^{2}}\left[1+\left(\frac{b}{r_{c}}\right)^{2}\right]^{-\frac{1}{2}}
\end{equation}Here 0.46 is the fraction of hydrogen atoms by number.

The baryonic gas fraction within galaxy clusters has recently received
attention \citep{WNE93,WF95}. Almost all the observations give
large values of baryon density $\Omega_{b}$ than expected  from the
theory of the big bang nucleosynthesis if $\Omega_{0} = 1$, which is
taken as an indication of a low density universe. Both local and high
redshift observations \citep{RFP99,EF99} suggest $f_{gas}$ scatters
between $0.1$ and $0.3$. To simplify, we use $f_{gas} = 0.2$
throughout the paper.

\subsection{Distribution Function}

In analogy to the Lyman-$\alpha$ forest system, we define
$\partial^{2}P/\partial N^{i}\partial z$ as the number of absorption
systems along the line of sight with a column density between $N^{i}$
and $N^{i}+dN^{i}$ per unit redshift. Here $N^{i} = N(X^{i})$ is the
column density of ion $X^{i}$. If we define $\Sigma$ as the cross
section of a galaxy group or cluster, the distribution function is then 
\begin{equation}
\frac{\partial^{2}P}{\partial N^{i}\partial z} = \int_{T_{X}} dT_{X}
\frac{dn}{dT_{X}} \frac{d\Sigma}{d N^{i}} \frac{d \ell}{dz}
\end{equation}
Here $dn/dT_{X}$ is the cluster number density distribution at
different cluster temperatures, which can be obtained by the
Press-Schechter function and the cluster mass-to-temperature
relationship. $\ell$ is the path length. We refer to \citet{PL98} for
detailed calculations.

\section{Computed Distribution}

\subsection{the X-ray Forest in Different Cosmological Models}

Given different cosmological models, the X-ray forest distribution
$\partial^{2}P/\partial N^{i}\partial z$ is determined through
equation (6) via the Press-Schechter function (equation (1)). Here we compare
the X-ray forest distribution under three cosmological models: a
standard cold dark matter model (SCDM); a low-density open CDM model
(OCDM) and a low-density $\Lambda$-dominated CDM
model. Table~\ref{tbl-1} shows all the model parameters.

\placetable{tbl-1}

Observationally resonance absorption lines will be produced by
transitions with the ions of the most abundant elements with the
largest oscillator strength. For specificity we choose the ions \ion{O}{8}, \ion{Si}{14} and \ion{Fe}{25}. Table~\ref{tbl-2} gives all the
parameters. \citet{MLA96} measured the
metal abundances in four rich clusters with {\sl ASCA}. We adopt their
metallicity because we concentrate only on high column densities
produced by rich clusters. The redshift evolution of cluster
metallicities can also play an important role here, for simplicity
we treat them as constants. We adopt the solar abundances from
\citet{AG89}. The atomic data are from
\citet{VVF96}. Figure~\ref{fig:ion} shows the
ionization fractions \citep{MMC98}, in which
\ion{O}{8}, \ion{Si}{14} and \ion{Fe}{25}  show three different peak temperatures
around $2 \times 10^{6}$, $10^{7}$ and $5 \times 10^{7}$,
respectively. These temperatures roughly correspond to the galaxy
cluster mass around $10^{12} M_{\odot}$, $10^{13} M_{\odot}$ and
$10^{14} M_{\odot}$, according to the mass-temperature relationship
(3).

\placetable{tbl-2} \placefigure{fig:ion}

To compute the Press-Schechter function we use the codes based on
\citet{EH97} for the calculation of power
spectrum and mass variance $\sigma(M)$. However, their codes are
normalized by COBE data, which is on a scale larger than the size of
galaxy clusters, so we renormalize the power spectrum based on the
observations of the cluster abundance in the local universe.

Figure~\ref{fig:distFuncFe} through figure~\ref{fig:distFuncOx} give
the distribution of three elements: \ion{Fe}{25}, \ion{Si}{14} and \ion{O}{8} under
three cosmological models at different redshifts. The most striking
feature is that in all three ions, the standard cold dark model shows
a more rapid evolution than the two other low-density models. This
trend in general agrees with the evolutionary scenarios predicted by
different cosmological models. In the SCDM model, the density
perturbation grows as $(1+z)^{-1}$,  most clusters are formed recently
and the cluster number density declines very quickly as we move to
high redshift. On the other hand, low-density models predict slower
evolution. Figure~\ref{fig:nT} shows the comoving cluster abundances
as a function of X-ray temperature at different redshift predicted by
the Press-Schechter distribution. Although LCDM also shows some
decline of cluster abundances at high temperature, SCDM show the
strongest evolution of number density at temperatures $\geq 5 \times\
10^{6}\ {\rm K}$. Among all three ions, the distribution of \ion{Fe}{25}
shows the biggest decline (over three decades between $z=0$ and
$z=3$). The reason is that \ion{Fe}{25} has the highest peak temperature
($\sim 5\times10^{7}\ $K) in figure~\ref{fig:ion}, which corresponds
to the biggest decline in the cluster abundances (figure~\ref{fig:nT}).

\placefigure{fig:distFuncFe} \placefigure{fig:distFuncSi}
\placefigure{fig:distFuncOx} \placefigure{fig:nT}

The second feature is that all three distributions show a rapid cutoff
between $10^{16} {\rm cm^{-2}}$ and $10^{17} {\rm cm^{-2}}$. This is mainly
because the maximum ion column density can not exceed the central
column density of the galaxy clusters. The central ion
column density is given by equation (5). For instance, the peak
temperature of \ion{O}{8} is around $2.4 \times 10^{6} $K
(figure~\ref{fig:ion}), which roughly corresponds to a central \ion{O}{8}
column density of  $6 \times 10^{16} {\rm cm^{-2}}$. This explains why
figure~\ref{fig:distFuncOx} shows a sharp cutoff around this column
density.

The third feature is that for all three ions, the SCDM model predicts
more absorbers along the line-of-sight than OCDM and LCDM. This is
because the spatial density of clusters is very sensitive to the
present matter density - $\Omega_{0}$ (equation (1)). $\Omega_{0}$
represents the overall amplitude of the density fluctuation and
changing of $\Omega_{0}$ will dramatically change the spatial density
of virialized objects, the clusters of galaxies. This can be a
method to determine $\Omega_{0}$ in the future from observations of
the X-ray forest.

\subsection{Monte Carlo Simulations} 

To obtain statistics on the distribution of X-ray absorption lines, we
carry out a series of Monte Carlo simulations. Assuming both absorber
column density and redshift are independent random variables
\citep{MJ90}, we define the Probability Density Functions (PDF) for
each variable, 

\begin{eqnarray}
f(z) & = & \frac{1}{A}
\int_{N^{i}_{\min}}^{N^{i}_{\max}}\frac{\partial^{2}P}{\partial
N^{i}\partial z}dN^{i} \\ g(N^{i}) & = & \frac{1}{A}
\int_{z_{\min}}^{z_{\max}}\frac{\partial^{2}P}{\partial N^{i}\partial
z}dz \nonumber
\end{eqnarray} Here $A$ is the total
absorber number, given by the integration of the distribution function
over both redshift and ion column density. We set a range for each
variable as $N^{i}\in \left[10^{12}, 10^{17}\right] {\rm cm^{-2}}$ and $z
\in \left[0, 3\right]$. Using these PDFs we obtain column densities
and redshifts for 10,000 randomly selected
lines-of-sight. Figures~\ref{fig:AVECOL} and ~\ref{fig:AVERED} show
the average cumulative distribution of the absorption line numbers
vs. column density and redshift, respectively. Although \ion{Fe}{25} is too
scarce to give any statistical information, both figures do reflect
the three features of the X-ray forest we noted before. In
figure~\ref{fig:AVERED} we include only the absorption lines with
column density higher than $10^{15} {\rm cm^{-2}}$ because this is the
lowest column density which is detectable by {\sl Constellation X}
(see the next section). In this figure we see that, compared to the
LCDM and OCDM models, SCDM presents a larger number of absorption
lines. For instance, SCDM shows that on average ten \ion{O}{8} absorption
lines up to $z=1$, where OCDM and LCDM give only four and three lines,
respectively.

\placefigure{fig:AVECOL} \placefigure{fig:AVERED}

Assuming a velocity dispersion of $b \sim 300\ {\rm km\ sec^{-1}}$, the line
optical depth is obtained by $\tau(\nu) = N^{i}\sigma(\nu)$. Here
$\sigma(\nu)$ is the absorption cross section at frequency $\nu$
\citep{S78}. The line-of-sight transmission is
defined as $D \equiv e^{-\tau}$. Figures~\ref{fig:Fe_Tran} to ~\ref{fig:Ox_Tran} show the transmission of \ion{Fe}{25}, Si XIX and \ion{O}{8}
under the three cosmological models, to $z=3$.

\placefigure{fig:Fe_Tran} 
\placefigure{fig:Si_Tran}
\placefigure{fig:Ox_Tran}

\section{Detectability}

Generally most X-ray absorption lines we discuss here are narrow,
unresolved lines. For a weak absorption line, the equivalent width
$W_{eq}$ is given by \citep{S78}
\begin{equation}
\frac{W_{eq}}{E} =
\frac{b}{c}\int_{-\infty}^{+\infty}\left\{1-\exp\left(-\tau_{0}e^{-x^{2}}\right)\right\}dx
\end{equation} Where $E$ is the line energy, $b$ and $c$ are velocity
dispersion and light speed. Here $\tau_{0}$ is the optical depth at
line center,  $\tau_{0} \equiv Ns\lambda/\pi^{1/2}b$, where $N$ is ion
column density, $\lambda$ is wavelength and $s$ is the Einstein
absorption coefficient.

The detectability of the X-ray forest is limited by the spectral
resolving power and effective area of the spectrometer. A weak,
unresolved resonance line has $W_{eq} \ll \Delta E$, where $\Delta E$
is the bin width, determined by the instrument resolving power R ($R
\equiv E/\Delta E$). Suppose we use an instrument with resolving power
$R$ to observe an resonance line around energy E (keV). The source
spectrum has a continuum intensity of $F_{X}$ (in units of photons
${\rm cm}^{-2}\ {\rm s}^{-1}\ {\rm keV}^{-1}$) around E. Given an
observing time T, the minimum detectable equivalent width for an
unresolved absorption line is
\begin{equation}
W_{eq} \gtrsim \left(S/N\right) \left(\frac{E}{A_{eff} \cdot R \cdot
    F_{X} \cdot T}\right)^{\frac{1}{2}}
\end{equation} Here $\left(S/N\right)$ is the signal-to-noise ratio
    and $A_{eff}$ is the effective area, and we assume negligible
    background. 

For illustration we choose a typical spectrum of an X-ray bright
quasar with photon index $\Gamma = 2.5$ and flux $1.0\times 10^{-11}\ 
{\rm ergs\ cm^{-2}sec^{-1}}$ between 0.1 and 2.4 keV. The Galactic column
density is $5.0\times 10^{20}\ {\rm cm^{-2}}$. Given this representative
spectrum we calculate the minimum detectable equivalent width and
column density of some ion species for a particular observation
time. In Table~\ref{tbl-3}, we list three instruments: {\sl Chandra}
LETG/HETG, {\sl XMM} RGS and {\sl Constellation-X}
Calorimeter/Gratings. Assuming the absorption ions are located at $z =
0.5$, the equivalent width is calculated based on a S/N of 3 and an
integration time of 100 ksec. Comparing this table with
figure~\ref{fig:AVECOL} and ~\ref{fig:AVERED}, we find that \ion{O}{8} ion
is the best candidate for all three instruments. Using XSPEC 10, we
simulate this ``representative'' spectrum plus \ion{O}{8} absorption lines
from one realization of the LCDM model (Figure~\ref{fig:Ox_Tran}) on
{\sl XMM} RGS. Several tens of quasars with $z \geq 1$ have such
spectrum or are even brighter, so we put the redshift of this quasar
at $z = 1$. The LCDM simulations give three absorption lines with $z
\leq 1$ and $N \geq 10^{16} {\rm cm^{-2}}$. Table~\ref{tbl-4} lists the
simulated line properties. Then we fit the simulated spectrum with a
model only containing Galactic absorption and a single power law. The
$\chi^{2}$ plot of figure~\ref{fig:sim} clearly shows three absorption
lines. Notice here line 2 and line 3 are only separated by
approximately $3.7$ eV.

\placetable{tbl-3} 
\placetable{tbl-4} 
\placefigure{fig:sim}

\section{Summary}

In this paper we use a semi-analytic method to investigate the X-ray
forest. Following previous work \citep{PL98} we calculate the X-ray
distribution function of \ion{O}{8}, \ion{Si}{14} and \ion{Fe}{25} under the Standard cold dark matter model and two
low-density models. \ion{Fe}{25} shows the more rapid evolution of SCDM,
compared to OCDM and LCDM; These trends are milder for \ion{Si}{14} and \ion{O}{8}. Using Monte-Carlo simulation, we investigate the average
distribution of the X-ray forest. We find SCDM model presents more
absorption lines than OCDM and LCDM, which eventually might yield a
method of determining $\Omega_{0}$. We also select a typical spectrum
of an X-ray bright, distant quasar to explore the detectability of the
X-ray forest. We find for all three telescopes, there are at least
several \ion{O}{8} absorption lines detectable by {\sl Chandra}, {\sl XMM}
and {\sl Constellation-X}. This result is consistent with \citet{PL98}
and Hellsten et al. (1998).

The X-ray forest distribution function depends on the Press-Schechter
distribution of galaxy clusters, the cluster mass-temperature
relationship and the gas distribution inside galaxy clusters (equation
(6)). In the following we discuss several important factors which
can affect the result.

The key element of the Press-Schechter function is the mass variance,
$\sigma(M)$. The present mass variance $\sigma(z = 0, M)$ is
calculated from the present power spectrum filtered through a top-hat
window, and normalized by the the mass variance at $8h^{-1}$ Mpc,
$\sigma_{8}$. The estimation of $\sigma_{8}$ involves fitting the
Press-Schechter function with the observed spatial number density of
local clusters \citet{HA91, WNE93, VL96, VL99}. In this evaluation a crucial
relationship is the mass-temperature relationship of equation
(3). Although there are both numerical and observational evidences
for this relationship, it has a well-known problem of the
``recent-formation approximation'': the clusters observed today formed
just before we observe them. To resolve this problem two different
methods \citep{LC93,S94} were proposed
to substitute the Press-Schechter function, both of which gave nearly
the same result \citep{VL96}. Based on the mergering-halo
formalism of \citet{LC93}, \citet{VD98} derived a new $M_{vir}-T_{X}$ relation and claimed that
equation (3) overestimated temperature  evolution and so the numbers
of high-z clusters. Their conclusion would affect the X-ray forest
distribution function of \ion{Fe}{25} discussed here, but not \ion{O}{8}. The
reason is that the exponential term of Press-Schechter function will
only become crucial at high temperature ($> 10^{7}$K), well above the
temperature of the ionization peak of \ion{O}{8} at $\sim 2 \times
10^{6}$K. However it would be important to investigate this effect
because it would largely decrease the possibility of detecting the
highly-ionized heavy metal absorptions lines, such as \ion{Fe}{25}.

Another important uncertainty which can affect the X-ray
distribution function is the metal abundance inside clusters. Using
{\sl ASCA}, \citet{MLA96} shows the mean abundances of O, Si
and Fe of four galaxy clusters are 0.48, 0.65 and 0.32 respectively,
which are close to the values used in this paper. However, there are a
few factors which can affect the metal distribution. First, the
metal abundances can depend on the mass of the cluster or
group. Recent observations by \citet{HMB99}
show at temperature above 1 keV, the metal abundances are roughly
$0.3\ M_{\odot}$, with little variance, while at temperature below 1
keV, the metal abundances drop very fast \citep[and
references there in]{R97, DMM99}. If it is 
real, this effect can dramatically drop the possibility of observing
Oxygen absorption lines because of its low peak temperature of
ionization. Another factor is the assumption in this paper of constant
metal abundances upto redshift as high as $z \sim 3$, although this is
important only for the richest systems which have significant column
density. No direct evidence shows a constant metal abundances beyond
$z \sim 0.3$ \citep{ML97}.

Further progress in this subject relies on both numerical simulations
and observations. Large-scale simulations on the X-ray clusters would
provide us more accurate information on the X-ray forest
distribution. On the observation side, with the launch of {\sl
Chandra} and {\sl XMM}, we would expect a few \ion{O}{7} or \ion{O}{8}
absorption lines by observing low and moderate redshift quasars. In
the future, {\sl Constellation-X} will provide us superior spectrum of
the X-ray forest, with the pioneer of probing the number and
distribution of objects in the universe.

\acknowledgments{We are grateful to Greg Bryan for many useful discussions
and helpful suggestion on calculating the power spectrum. We thank
Rasalba Perna for kindly discussions on her paper. The code for
calculating power spectrum is provided by Daniel Eisenstein and Wayne
Hu. We would also like to thank MIT/CXC team for help on using X-ray
softwares. We also thank the suggestions from the referee. This work
is supported in part by NASA contract NAS 8-38249.}

\clearpage


\clearpage


\figcaption[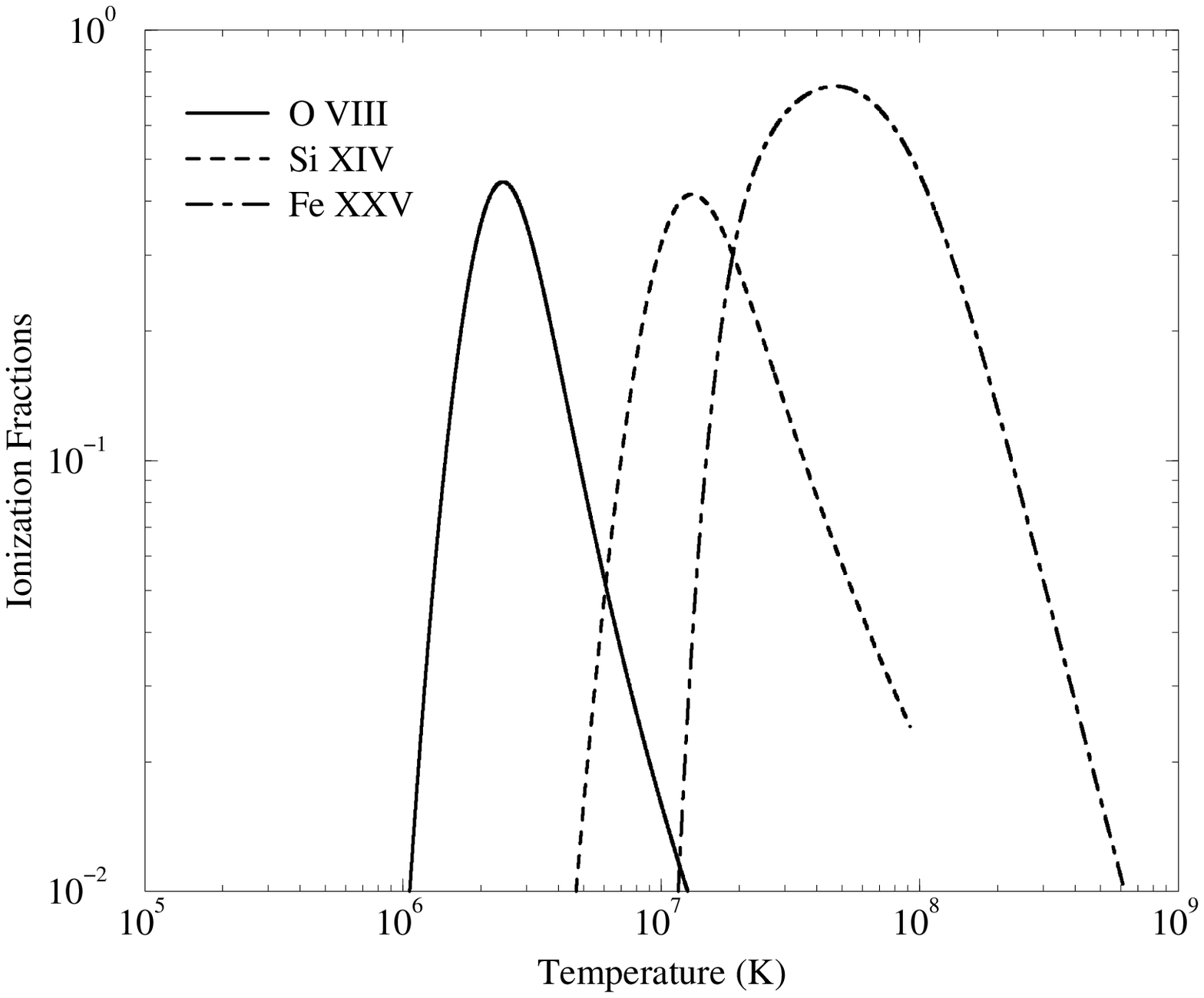]{Ionization fractions for \ion{O}{8} (solid curve),
\ion{Si}{14} (dashed curve) and \ion{Fe}{25} (dot-dashed curve).
\label{fig:ion}}

\figcaption[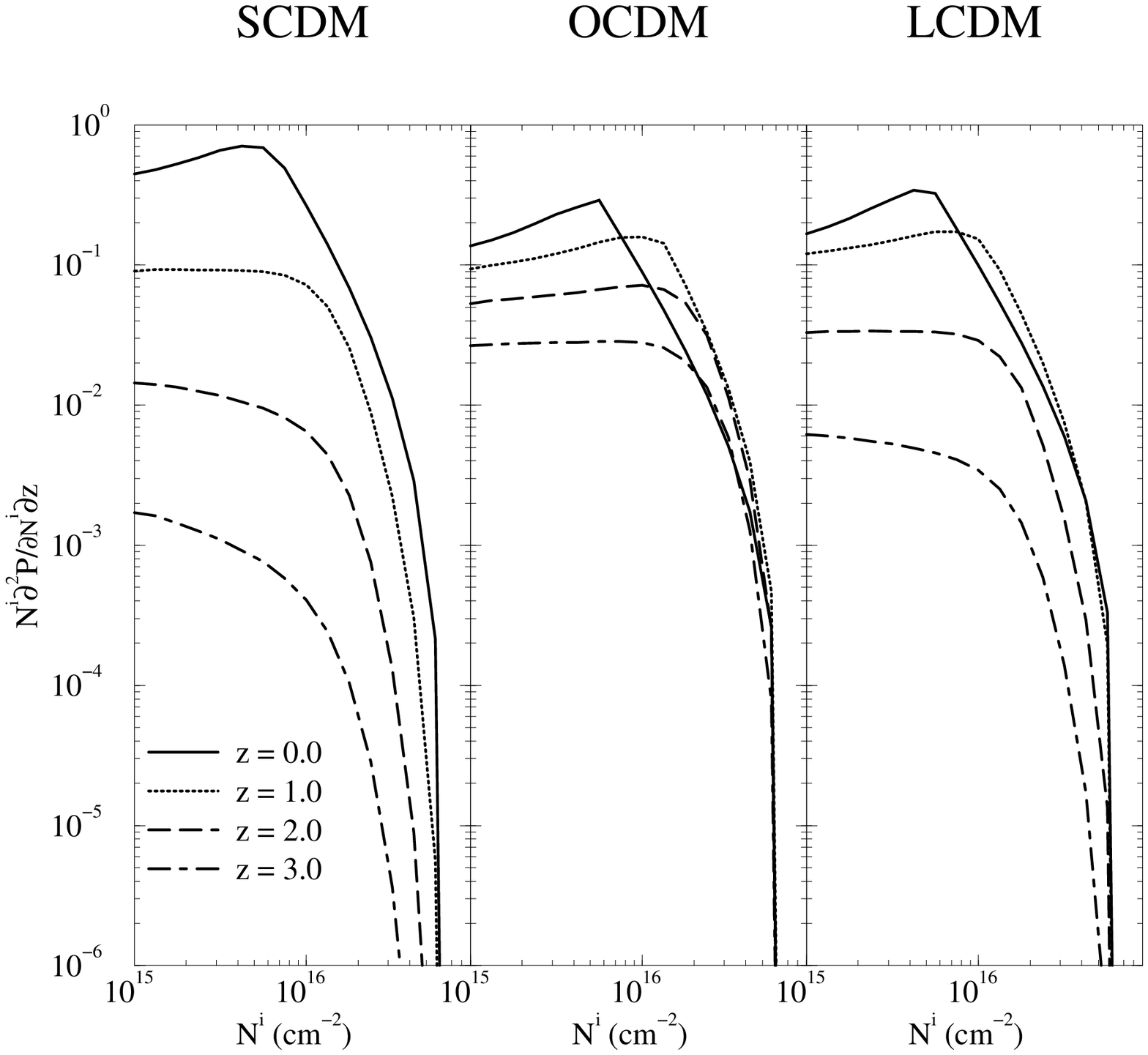]{The distribution functions of number of
absorbers per redshift and column density for \ion{Fe}{25}. For each
of the three cosmological models, the distribution function vs. column
density is given at four different redshifts from top to bottom : $z=0.0$(solid
line), $z=1.0$(dotted line), $z=2.0$(dashed line), $z=3.0$(dot-dashed
line). There is little evolution in the $\Omega_{0} < 1$ cosmological
models, comparing to a more rapid evolution in the $\Omega_{0} = 1$
model. \label{fig:distFuncFe}}

\figcaption[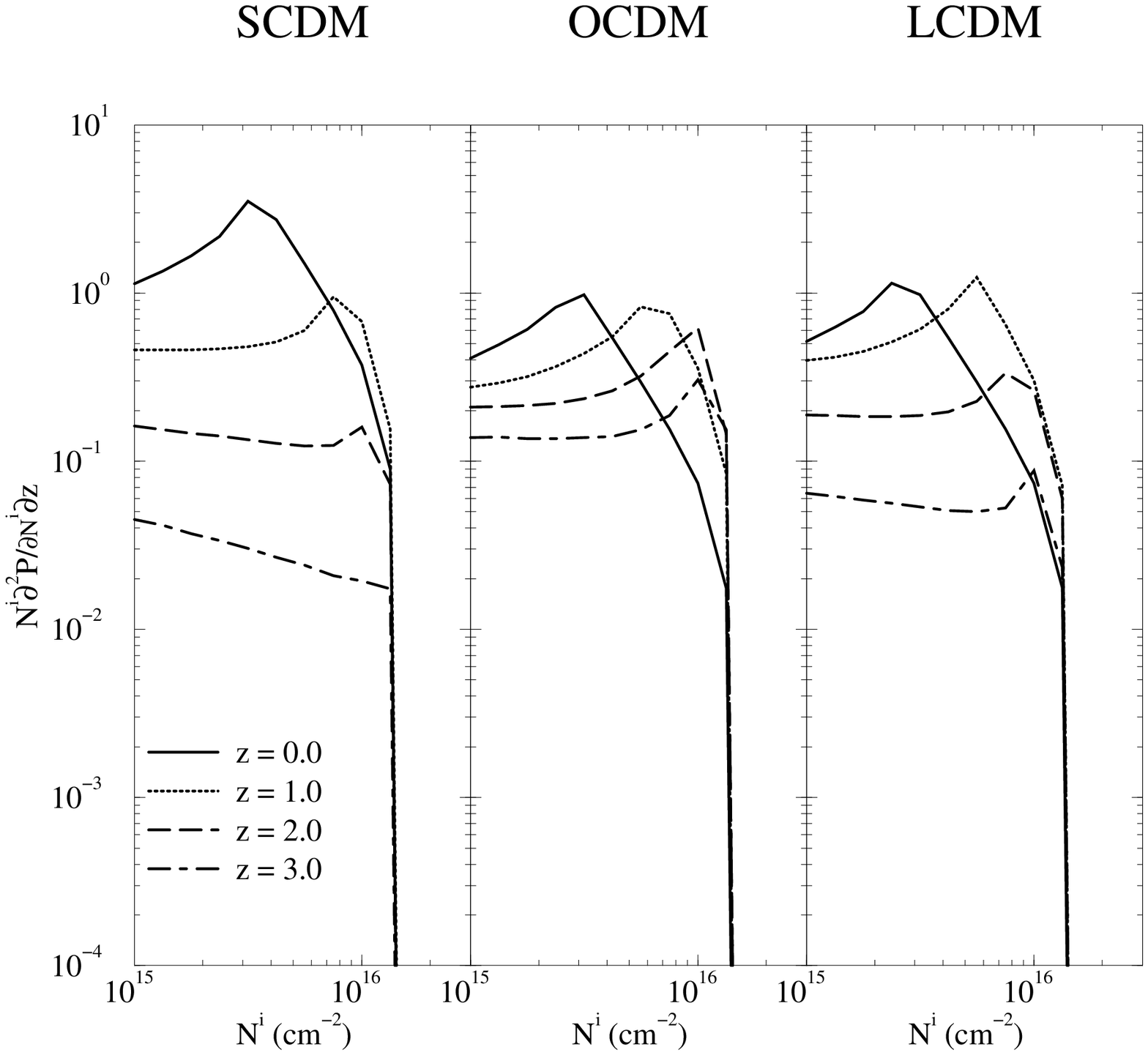]{The distribution function for \ion{Si}{14}. Line symbols are
the same as the previous figure. The evolution of $\Omega_{0} = 1$
model is still faster than low density models, but is milder than Fe
XXV. \label{fig:distFuncSi}}

\figcaption[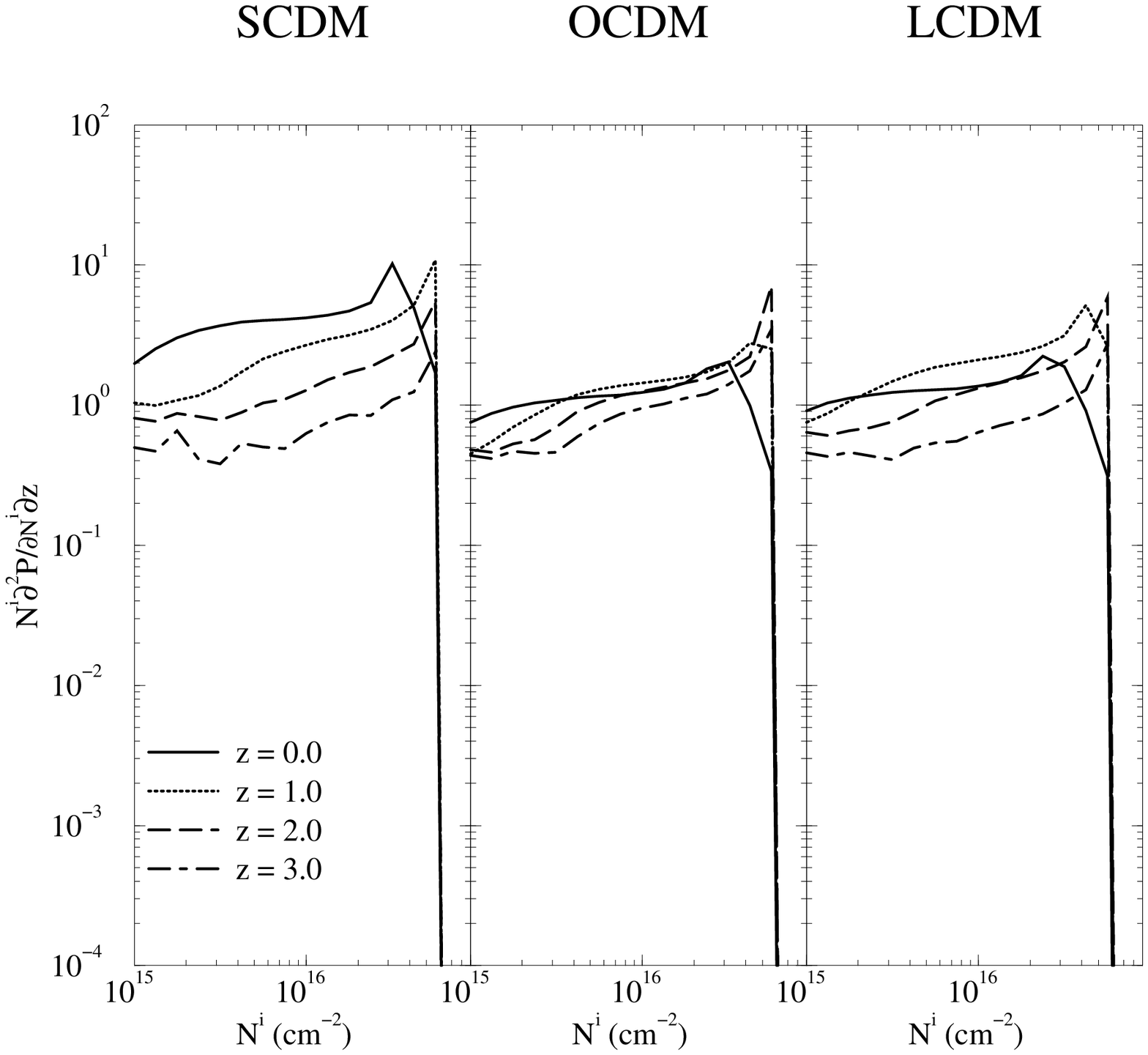]{The distribution function for
\ion{O}{8}. Line symbols are the same as the previous
figure. \label{fig:distFuncOx}} 

\figcaption[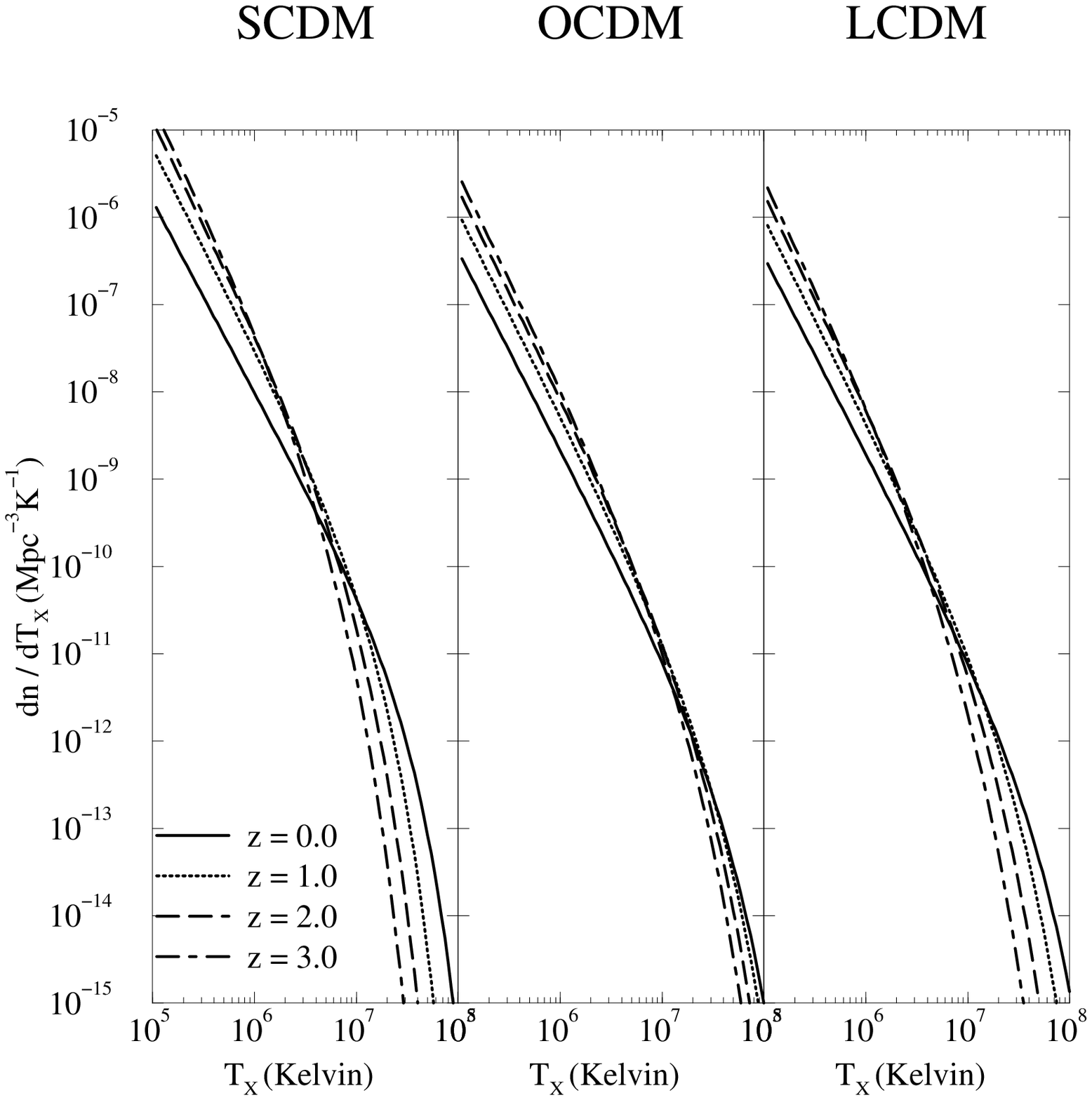]{Temperature distribution of galaxy clusters at different
redshifts, predicted by the Press-Schechter function. For each of the
three cosmological models, temperature distribution are plotted from
$z=0$ to $z=3$. SCDM shows a rapid evolution at high temperature,
compared to OCDM and LCDM. \label{fig:nT}}

\figcaption[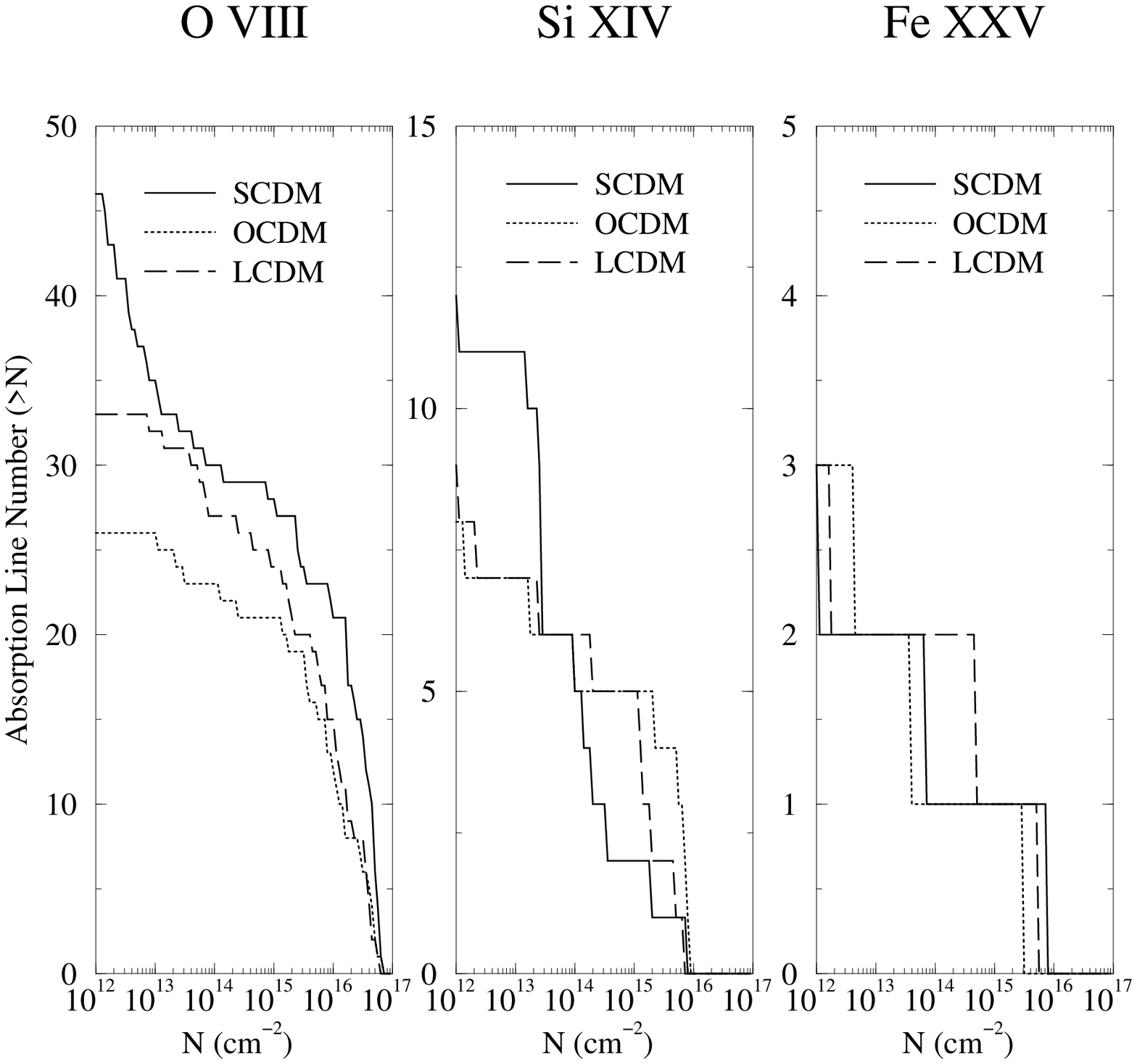]{Cumulative column density distribution of average absorption
line number upto $z=3$. Each plot gives the absorption line numbers
for one ion under the three cosmological models. \ion{O}{8} shows more
absorption lines. For instance, in the SCDM model, \ion{O}{8} gives
twenty-nine lines with column density over $10^{15} {\rm cm^{-2}}$, compared
to five for \ion{Si}{14} and one for \ion{Fe}{25}. \label{fig:AVECOL}}

\clearpage

\figcaption[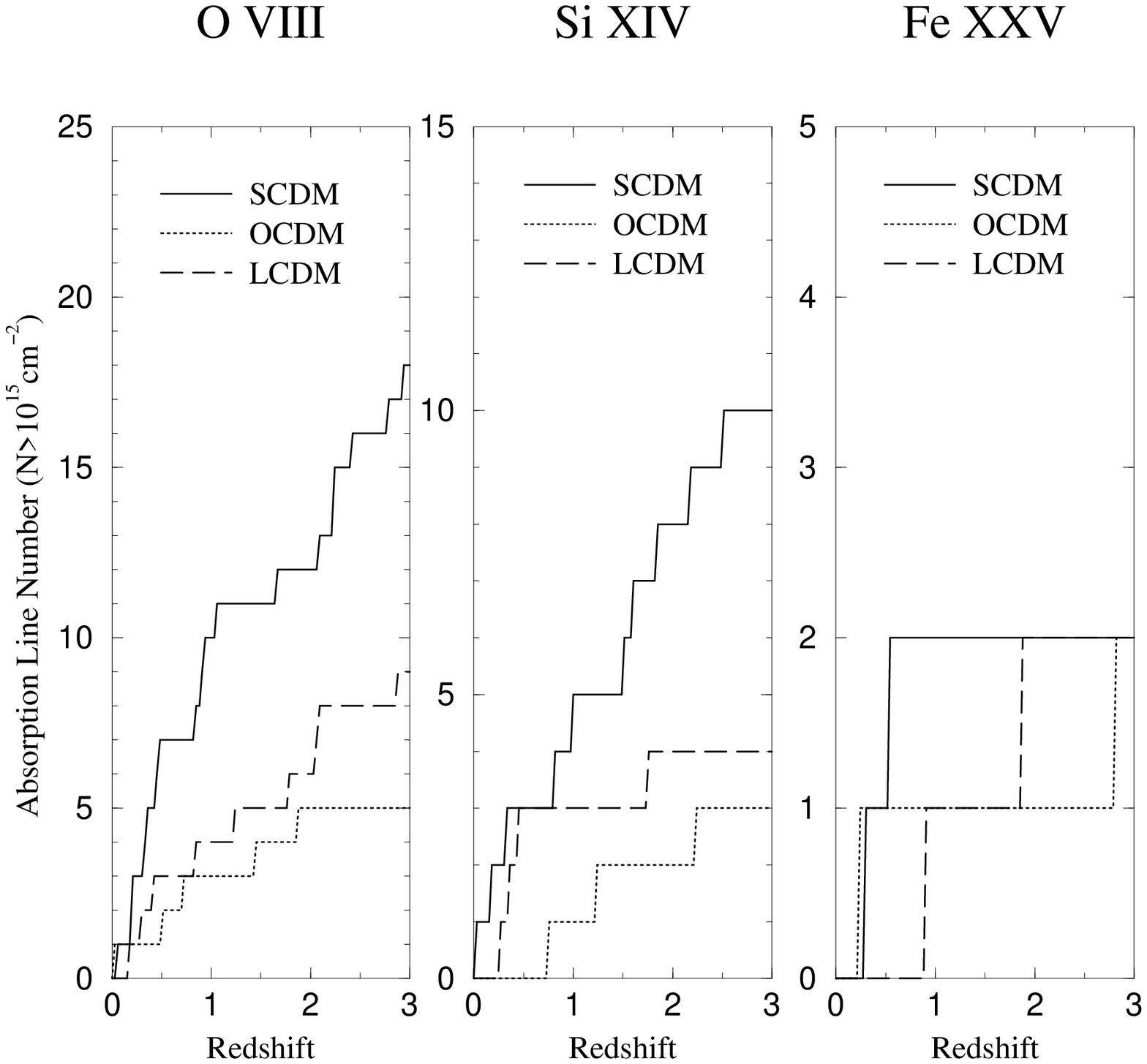]{Cumulative redshift distribution of average absorption line
number upto $z=3$. Each plot shows the absorption line number with
column density higher than $10^{15} {\rm cm^{-2}}$ under the three
cosmological models. In general SCDM shows more rapid
evolution. \label{fig:AVERED}}

\figcaption[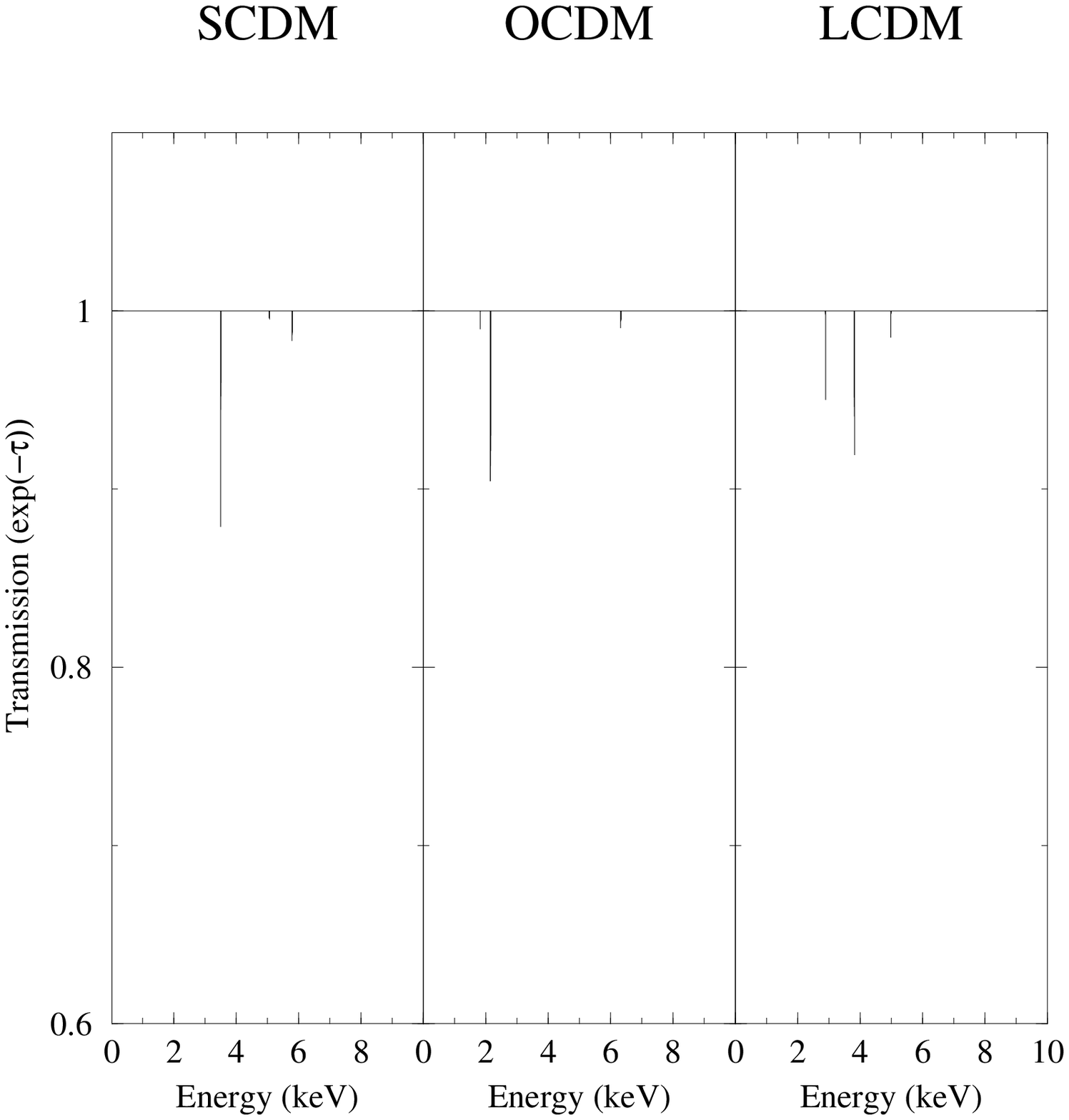]{Transmission of \ion{Fe}{25} under the three cosmological
models. The line energy is 6.70 keV and spreads over redshift space
from $z=0$ to $z=3$. \label{fig:Fe_Tran}}

\figcaption[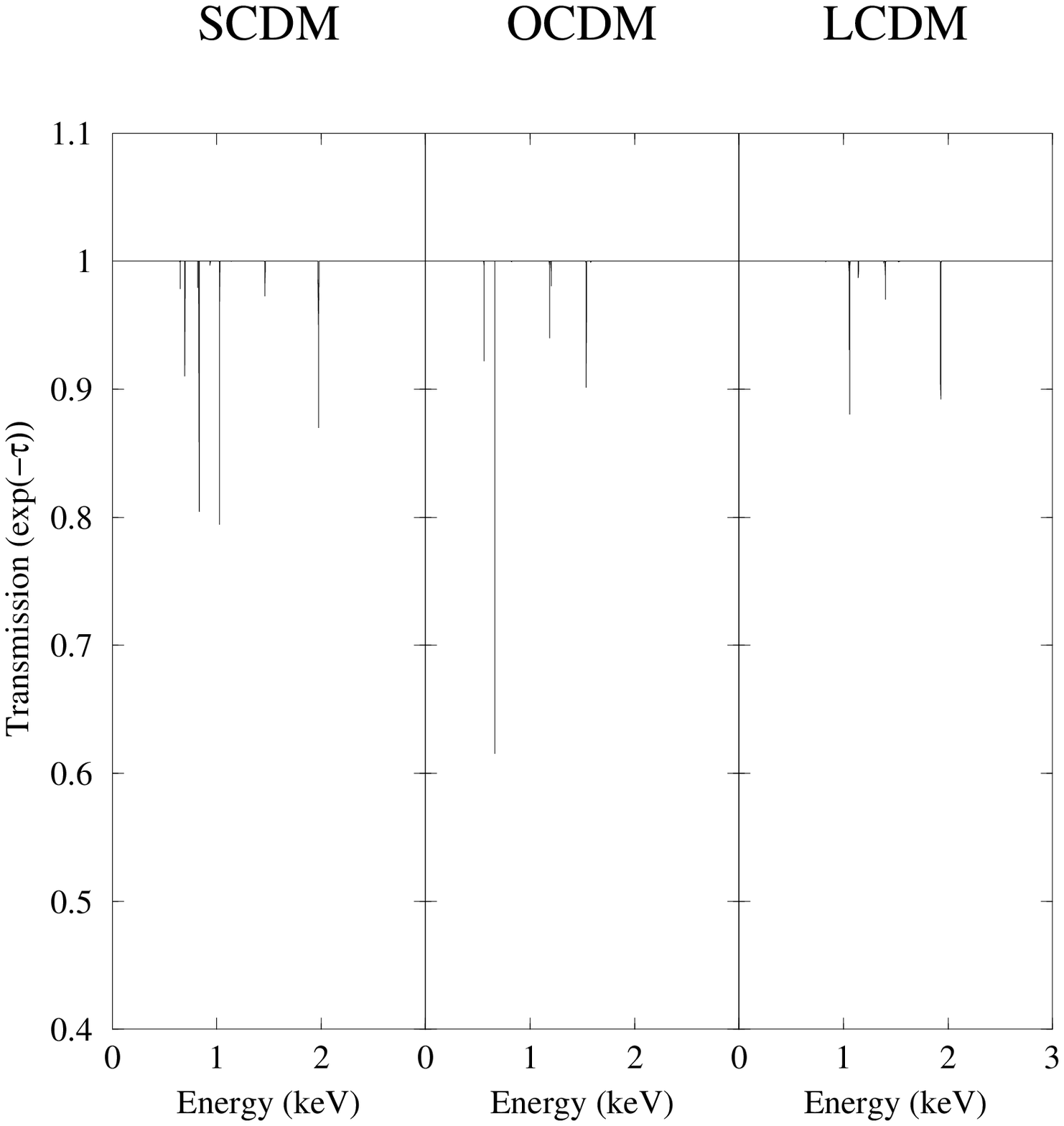]{Transmission of \ion{Si}{14} under the three cosmological
models. The line energy is 2.01 keV and spreads over redshift space
from $z=0$ to $z=3$. \label{fig:Si_Tran}}  

\figcaption[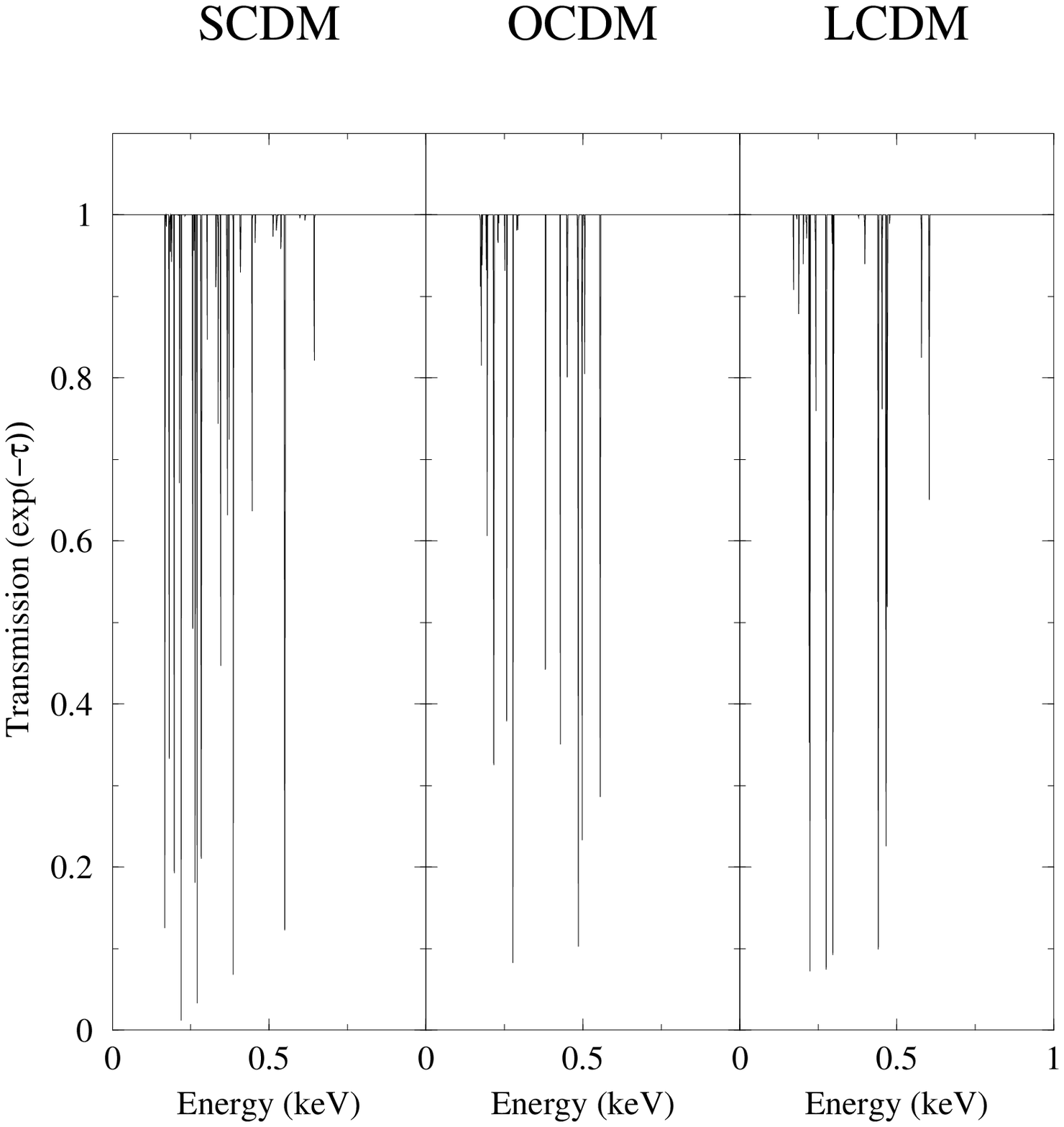]{Transmission of \ion{O}{8} under the three cosmological
models. The line energy is 0.65 keV and spreads over redshift space
from $z=0$ to $z=3$. \label{fig:Ox_Tran}}  

\figcaption[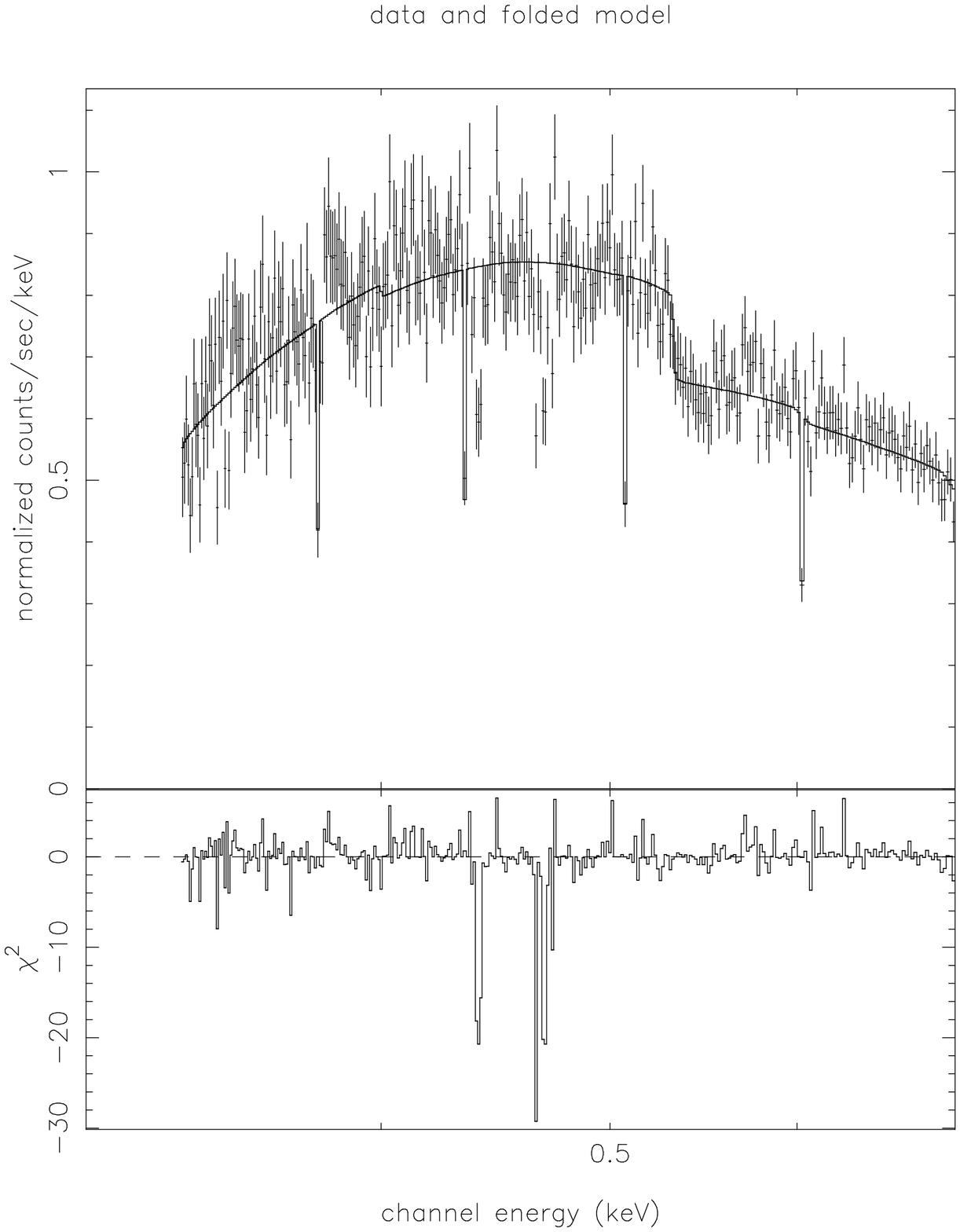]{The upper plot is the simulated spectrum,
containing three \ion{O}{8} absorption line at 440.11 eV, 464.82 eV and 468.48 eV,
respectively. We fit the simulated spectrum by a simple powerlaw plus
the Galactic absorption. The lower plot shows the $\chi^{2}$, which
clearly indicates three absorption features at the corresponding
energy. The four big gaps in the upper plot are due to the instrumental
gaps. \label{fig:sim}}  

\clearpage


\begin{deluxetable}{cccccccc}
\tablewidth{0pt} \tablecaption{The Cosmological Models \label{tbl-1}}
\tablehead{ \colhead{Model} & \colhead{$\Omega_{0}$} &
\colhead{$\Omega_{R}$} & \colhead{$\Omega_{\Lambda}$} &
\colhead{$h$\tablenotemark{a}} & \colhead{$n$\tablenotemark{b}} &
\colhead{$\sigma_{8}$}   } \startdata SCDM & 1.0 & 0.0 & 0.0 & 0.7 &
1.0 & 0.52\nl OCDM & 0.3 & 0.7 & 0.0 & 0.7 & 1.0 & 0.87\nl LCDM & 0.3
& 0.0 & 0.7 & 0.7 & 1.0 & 0.93\nl \enddata \tablenotetext{a}{Here we
set $H_{0} = 70 $ km ${\rm s}^{-1} {\rm Mpc}^{-1}$}
\tablenotetext{b}{We choose a scale-invariant power spectrum with
index $n=1$}

\end{deluxetable}

\begin{deluxetable}{ccccc}
\tablewidth{0pt} \tablecaption{Ion Transition Parameters
\label{tbl-2}} \tablehead{ \colhead{Ion} &  \colhead{Energy (keV)} &
\colhead{Oscillator Strength} & \colhead{Abundance\tablenotemark{a}} &
\colhead{Solar Abundance\tablenotemark{b}}  } \startdata \ion{O}{8} &
0.65 & 0.416 & 0.5 & $8.53 \times 10^{-4}$\nl \ion{Si}{14} & 2.01 &
0.416 & 0.5 & $3.58 \times 10^{-5}$\nl \ion{Fe}{25} & 6.70 & 0.798 &
0.3 & $3.23 \times 10^{-5}$\nl \enddata \tablenotetext{a}{Relative to
solar abundance $Z_{\odot}$.}  \tablenotetext{b}{Relative to hydrogen
number density.}
\end{deluxetable}

\begin{deluxetable}{cccc}
\tabletypesize{\footnotesize} \tablewidth{0pt} \tablecaption{Resonant
X-ray Absorption Lines at $z=0.5$ \label{tbl-3}} \tablehead{
\colhead{Ion} & \colhead{Instrument} &  \colhead{Equivalent Width
(eV)} &  \colhead{Column Density (${\rm cm^{-2}}$)}}  \startdata
\ion{O}{8} & {\sl Chandra} (LETG)               & 0.69  & $3.5\ 10^{16}$ \\
& {\sl XMM} (RGS)                    & 0.21  & $7.6\ 10^{15}$ \\ &
{\sl Constellation X} (Gratings)   & 0.04  & $1.3\ 10^{15}$\\ \cline{1-4}
\ion{Si}{14} & {\sl Chandra} (MEG)                & 1.34  & $4.7\ 10^{16}$
\\ & {\sl XMM} (RGS)                    & 2.25  & $1.1\ 10^{17}$ \\ &
{\sl Constellation X} (Calorimeter)& 0.09  & $3.1\ 10^{15}$\\ \cline{1-4}
\ion{Fe}{25} & {\sl Chandra} (HEG)                & 24.7& $1.7\ 10^{20}$ \\
& {\sl XMM} (RGS)                    & N/A\tablenotemark{a} &
N/A\tablenotemark{a} \\ & {\sl Constellation X} (Calorimeter)& 0.52 & $9.2\
10^{15}$\\ \enddata \tablenotetext{a}{The energy range of XMM RGS is
$0.35-2.5 keV$}
\end{deluxetable}

\begin{deluxetable}{ccccc}
\tablecaption{Selected \ion{O}{8} Absorption Lines for XMM Simulation
\label{tbl-4}}
\tablehead{ \colhead{line number} & \colhead{Column Density
(${\rm cm^{-2}}$)} & \colhead{Redshift} & \colhead{Energy (eV)} &
\colhead{Equivalent Width (eV)}} \startdata 1 & $4.28\ 10^{16}$ &
0.486 & 440.11 & 0.54 \\ 2 & $2.80\ 10^{16}$ & 0.396 & 468.48 & 0.80
\\ 3 & $3.27\ 10^{16}$ & 0.407 & 464.82 & 0.79 \\ \enddata
\end{deluxetable}

\clearpage

\end{document}